\documentclass{amsart}

\usepackage{epsfig}
\usepackage{graphicx}
\usepackage{amsmath,amssymb,amsfonts,latexsym}
\usepackage{graphicx}

\usepackage{amssymb}
\usepackage{enumerate, xspace}
\usepackage{dsfont}
\usepackage{afterpage}
\usepackage{changebar}
\usepackage{amsthm}
\usepackage{rotating}

\numberwithin{equation}{section}

\begin{document}

\title[Resistance in quantum Hall metals]
{Hall resistance in quantum Hall metals due to Pancharatnam phase retardation and \\energy level spacing}
\author {Andrew Das Arulsamy}
\address{Condensed Matter Group, Division of Interdisciplinary Science, F-02-08 Ketumbar Hill, Jalan Ketumbar, 56100 Kuala-Lumpur, Malaysia}
\email{sadwerdna@gmail.com}


\date{\today}

\begin{abstract}
We derive the trial Hall resistance formula for the quantum Hall metals to address both the integer and fractional quantum Hall effects. Within the degenerate Landau levels, Zeeman splitting and level crossings in the presence of changing magnetic-field strength determine the Pancharatnam phase retardation, including the phase acceleration or deceleration, which are related to the changes in the phase and group momenta of a wavefunction. We discuss the relevant physical postulates with respect to Pancharatnam phase retardation to qualitatively reproduce the measured Hall resistance's zigzag curve for both the integer and the fractional filling factors. Along the way, we give out some hints to falsify our postulates with experiments.  
\end{abstract}

\maketitle

\section{Introduction}

Since the discovery of the electron as a negatively charged sub-atomic particle by Thomson~\cite{thom}, its spin by Uhlenbeck and Goudsmit~\cite{goud}, and its wave nature by de Broglie~\cite{de}, we have always imagined (in one way or another) of their interactions and motion within a given atom or an electronic device, and how their wavefunctions transform in order to flow along the grid and through the lattice points, satisfying the principle of least action~\cite{mor} (when the system is away from the critical point) and the principle of maximum interaction~\cite{jcs} during a phase transition. This single thought have led us to formulate the proper and consistent theories to understand Fermi liquid metals, semiconductors, band insulators, conventional superconductors, magnets, ferromagnets, ferroelectrics and the Mott-Hubbard insulators~\cite{ash}. Apart from the high temperature superconductors~\cite{bed} and strange metals~\cite{ander}, topological insulators~\cite{hat,kane}, and quantum Hall metals~\cite{klit,tsui,laugh,laugh2} are the other major solid-state systems that currently need proper formulations to understand their electronic properties, regardless whether these systems have the potential for future technological marvels.     

Here, we will derive the trial Hall resistance formula that will capture the essential physics required to understand the constant Hall resistances for certain filling factors and the magnetic-field dependent part of the Hall resistance due to the finite longitudinal resistance along the applied electric fields. The physical mechanisms invoked to formulate the Hall resistance formula involve the usual metallic Fermi liquid where the degenerate Landau levels (LLs) are crossed, and at each point of this LL crossing, there exists a finite but irrelevant energy-level spacing, which is responsible for the nonzero longitudinal resistance along the applied electric fields with increasing magnetic fields. The above energy-level spacings are called irrelevant because they do not require the wavefunction transformation beyond the phase factor, otherwise, these spacings are known as the ``relevant'' energy-level spacings, which have been proven elsewhere~\cite{ptp}. 

Anyway, these irrelevant LL spacings do not play any role for certain LL crossings, hence can be taken to be zero symbolically for convenience. The system's LLs split due to Zeeman splitting, and with increasing magnetic field strength, these levels cross each other in a complicated way (within the degenerate LLs), which can be used to postulate two independent physical processes. The first physical mechanism is influenced by this irrelevant LL spacings such the electron-flow requires large changes to the phase factor and group momentum of a wavefunction, while the other mechanism for electron-flow is independent of these spacings (require small changes to the phase factor, or negligible changes to the group momentum of a wavefunction). Here, we will further elaborate on these two processes, and make use of them to derive and discuss the Hall resistance in quantum Hall metals for both fractional and integer filling factors. 

\section{Integer Hall conductance}

We start with these open sets, $U_j$ and $U_k$, which are allowed to overlap such that $U_j \cap U_k \neq \emptyset$ and $\{U_j,U_k\} \in M^m_{\mathbb{C}}$ where $M^m_{\mathbb{C}}$ is a complex $m$-dimensional topological space and an $m$-dimensional differential manifold that deals with complex ($\mathbb{C}$) numbers. We also define a transition function,
\begin {eqnarray}
\texttt{t}_{jk}(\texttt{p}) = \exp[{\rm i}\gamma(\texttt{p})], \label{eq:1}
\end {eqnarray}  
that smoothly maps $U_j$ to $U_k$, following Definition 5.1 and Theorem 10.2 given by Nakahara in Ref.~\cite{naka} where i = $\sqrt{-1}$, $\gamma(\texttt{p}) \in \mathbb{R}$, an arbitrary function that will be defined later as the Pancharatnam-Berry phase, and $\texttt{p}$ refers to the same coordinate point on both $U_j$ and $U_k$. Physically, we can define $U_j = \{\psi_{\rm VB}(\textbf{k})_J\}$ and $U_k = \{\psi_{\rm CB}(\textbf{k})_K\}$ where $\textbf{k}$ is the wave vector in momentum space, $|\textbf{k}|$ denotes the wave number, $j = 1$, $k = 2$, $\{J,K\} \in \mathbb{N}^*$ (excluding zero), $\psi_{\rm VB}(\textbf{k})_J$ and $\psi_{\rm CB}(\textbf{k})_K$ are the electron wavefunctions in the valence and conduction energy bands, respectively, and therefore $J$ and $K$ are the principal quantum numbers such that $J < K$. Moreover, $U_j \cap U_k \neq \emptyset$ implies the energy gap, $E_{\rm gap} = 0$ and therefore, $\psi_{\rm CB}(\textbf{k})_K = \psi_{\rm VB}(\textbf{k})_J = \psi_J(\textbf{k})$. If $U_j \cap U_k = \emptyset$, then $E_{\rm gap} \neq 0$ and consequently $\texttt{t}_{jk}(\texttt{p})$ (defined in Eq.~(\ref{eq:1})) does not exist because $\texttt{t}_{jk}(\texttt{p})$ strictly requires zero energy gap, and accordingly, we now define the respective Pancharatnam-Berry (PB) connection and PB curvature~\cite{naka}, 
\begin {eqnarray}
\mathcal{A}_k(\texttt{p}) = \texttt{t}_{jk}(\texttt{p})^{-1}\mathcal{A}_j(\texttt{p})\texttt{t}_{jk}(\texttt{p}) + \texttt{t}_{jk}(\texttt{p})^{-1}\textbf{d}\texttt{t}_{jk}(\texttt{p}), \label{eq:2}
\end {eqnarray} 
\begin {eqnarray}
\mathcal{F}_{j}(\texttt{p}) = \textbf{d}\mathcal{A}_{j}(\texttt{p}) + \mathcal{A}_{j}(\texttt{p}) \wedge \mathcal{A}_{j}(\texttt{p}). \label{eq:3}
\end {eqnarray} 
Here $\textbf{d}$ = $[\partial/\partial \texttt{x}]{\rm d}\texttt{x}$ ($\texttt{x}$ here can be any variable) and $\wedge$ denotes the exterior product, whereas $\textbf{d}$ and d are the exterior derivatives. We now drop $\texttt{p}$ for convenience, which will become clear shortly. Furthermore, the component of a PB connection is given by $\mathcal{A}_{j\mu} = [\partial\mathcal{A}_{j\mu}/\partial x^{\mu}]$d$x^{\mu}$ where $1 \leq \mu < \nu \leq m$. The origin of Berry's phase~\cite{ber2} has been proven earlier by Pancharatnam using the elliptically-polarized pencil beams within the Poincar${\rm \acute{e}}$ sphere~\cite{panch,panch2,panch3}, such that, the Berry's phase is a rediscovered special case (with a constant group momentum) within the Pancharatnam's wavefunction transformation, which has been proven in Ref.~\cite{panch3} using the generalized theory of interference put forth by Pancharatnam~\cite{panch,panch2}. The Pancharatnam wavefunction transformation occurs whenever a wavefunction picks up, or drops a phase factor~\cite{panch3}.

The two-dimensional quantized conductance can be obtained by first defining the respective PB connection and curvature,
\begin {eqnarray}
\mathcal{A}_{j(\mu,\nu)} = \langle\psi_J| \wedge \bigg(\frac{\partial|\psi_J\rangle}{\partial x^{\mu,\nu}}{\rm d}x^{\mu,\nu}\bigg), \label{eq:10}
\end {eqnarray}
\begin {eqnarray}
\mathcal{F}_{j(\mu\nu)} &=& \nabla \times \mathcal{A}_{j(\mu,\nu)} \nonumber \\&=& \bigg[\bigg(\frac{\partial\langle\psi_J|}{\partial x^{\mu}}\bigg)\bigg(\frac{\partial|\psi_J\rangle}{\partial x^{\nu}}\bigg) - \bigg(\frac{\partial\langle\psi_J|}{\partial x^{\nu}}\bigg)\bigg(\frac{\partial|\psi_J\rangle}{\partial x^{\mu}}\bigg)\bigg]{\rm d}x^{\mu} \wedge {\rm d}x^{\nu}. \label{eq:11}
\end {eqnarray} 
Here, we have made use of these equalities, ${\rm d}x^{\mu} \wedge {\rm d}x^{\nu} = {\rm d}x^{\nu} \wedge {\rm d}x^{\mu}$, $\partial^2/\partial x^{\mu}\partial x^{\nu} = \partial^2/\partial x^{\nu}\partial x^{\mu}$ and the assumption $\mathcal{F}_{j(\mu\nu)} = \nabla \times \mathcal{A}_{j(\mu,\nu)}$. Moreover, the subscripts $j$ and $k$ now refer to $U_j \cap U_k = U'_j = U'_k = U$ where we have confined the relevant coordinate point ($\texttt{p}$) and functions ($\texttt{t}_{jk}$ and $\psi_J$) within the overlapped open set, $U \in M^m_{\mathbb{C}}$ because $\texttt{t}_{jk}$ does not exist if $\psi_J \notin U$. We also have made use of the PB phase~\cite{panch3,david},   
\begin {eqnarray}
\gamma_J(T) = {\rm i}\oint\langle\psi_J(\textbf{X})|\nabla\psi_J(\textbf{X})\rangle {\rm d}\textbf{X}, \label{eq:12}
\end {eqnarray}  
to obtain the definition given in Eq.~(\ref{eq:10}), and this definition is also exactly identical to the one used by Nakahara~\cite{naka}. In Eq.~(\ref{eq:12}), $T$ is the time taken for the Hamiltonian to return to its original form. Obviously, Eq.~(\ref{eq:12}) is nonintegrable and cannot be zero if $\psi_J(\textbf{k}) \rightarrow \psi_J[X_1(\textbf{k}),X_2(\textbf{k}),\cdots,X_N(\textbf{k})] = \psi_J(\textbf{X})$, $N > 1$ and at least $X_1(\textbf{k}) \neq X_2(\textbf{k})$. Using Eq.~(\ref{eq:11}) and the Stokes theorem,
\begin {eqnarray}
\oint_{\texttt{C}} \mathcal{A}_{j(\mu,\nu)} = \int_{\texttt{S}}\mathcal{F}_{j(\mu\nu)}, \label{eq:13}
\end {eqnarray}  
we can now rewrite Eq.~(\ref{eq:12}),   
\begin {eqnarray}
\gamma_J(nT) = {\rm i}\int_{\texttt{S}}\big[\nabla \times \mathcal{A}_{j(\mu,\nu)}\big] = n2\pi, \label{eq:14}
\end {eqnarray}  
where $\texttt{C}$ and $\texttt{S}$ refer to the curve (that forms a closed loop) and the surface integrals, respectively, in the momentum ($\textbf{k}$) space, while $n \in \mathbb{Z}^*$ is related to the Chern winding number, and $\mathbb{Z}^*$ is the set of positive integers, including zero. Obviously, $n$ has got to be an integer due to Eq.~(\ref{eq:1}) if $\exp{({\rm i}\gamma(\texttt{p}))} = 1$ where $\gamma(\texttt{p}) = [0,n2\pi]$. 

We now prove $\mathcal{F}_{j(\mu\nu)} - \mathcal{F}_{k(\mu\nu)} = \textbf{d}_{\mu}\mathcal{A}_{j\nu} - \textbf{d}_{\nu}\mathcal{A}_{k\mu} = \nabla \times \mathcal{A}_{j(\mu,\nu)}$ by using the fact that $\mathcal{F}_{j(\mu\nu)} = \textbf{d}\mathcal{A}_{j(\mu,\nu)}$, which is a special case of Eq.~(\ref{eq:3}) that has been proven to be exact by Nakahara using the Poincar${\rm \acute{e}}$'s lemma~\cite{naka}. 
\begin{proof} Recall Eq.~(\ref{eq:2}), and using Eq.~(\ref{eq:1}), one obtains
\begin {eqnarray}
\mathcal{A}_{k(\mu,\nu)} = \mathcal{A}_{j(\mu,\nu)} + {\rm id}\gamma. \label{eq:15}
\end {eqnarray}
Consequently,  
\begin {eqnarray}
\gamma &=& {\rm i}\oint_{\texttt{C}}\big(\mathcal{A}_{j\nu} - \mathcal{A}_{k\mu}\big) \label{eq:16a} \\&=& {\rm i}\int_{\texttt{S}}\bigg[\bigg(\frac{\partial\langle\psi_J|}{\partial x^{\mu}}\bigg)\bigg(\frac{\partial|\psi_J\rangle}{\partial x^{\nu}}\bigg) - \bigg(\frac{\partial\langle\psi_J|}{\partial x^{\nu}}\bigg)\bigg(\frac{\partial|\psi_J\rangle}{\partial x^{\mu}}\bigg)\bigg]{\rm d}x^{\mu} \wedge {\rm d}x^{\nu} \label{eq:16} \\&=& n2\pi. \label{eq:17}
\end {eqnarray}  
To arrive at Eq.~(\ref{eq:16}), we also have made use of the equalities introduced after Eq.~(\ref{eq:11}), the Stokes theorem (Eq.~(\ref{eq:13})) and the fact that 
\begin {eqnarray}
\mathcal{A}_{k(\mu,\nu)} = \langle\psi_J|\exp{(-{\rm i}\gamma)} \wedge \bigg(\frac{\partial|\psi_J\rangle}{\partial x^{\mu,\nu}}{\rm d}x^{\mu,\nu}\bigg)\exp{({\rm i}\gamma)}, \label{eq:10a}
\end {eqnarray}
due to Eq.~(\ref{eq:1}). The integrand in Eq.~(\ref{eq:16}) is nothing but what is given in Eq.~(\ref{eq:11}).
\end{proof} 

For a metallic free-electron or Fermi liquid system, the definition for the Hall conductance is given by $G_{\rm Hall} = (e^2/h)n_{\rm Chern}$ where $e$ denotes the electron charge, $h$ is the Planck constant and the integer $n_{\rm Chern}$ is the Chern winding number that can be obtained from Eq.~(\ref{eq:14}) or Eq.~(\ref{eq:16}) such that the Chern winding number is given by 
\begin {eqnarray}
n = n_{\rm Chern} &=& \frac{{\rm i}}{2\pi}\int_{\texttt{S}}\big[\nabla \times \mathcal{A}_{j(\mu,\nu)}\big] \label{eq:18} \\&=& \frac{{\rm i}}{2\pi}\int_{\texttt{S}}\big[\mathcal{F}_{j(\mu\nu)} - \mathcal{F}_{k(\mu\nu)}\big], \label{eq:18a}
\end {eqnarray}  
following the definition, $n_{\rm Chern} = c_1(\mathcal{F}) = {\rm i}\mathcal{F}/2\pi$ given in Ref.~\cite{naka} within the first Chern class ($c_1(\mathcal{F})$) where $\mathcal{F}$ is arbitrarily known as the field strength, which can be related to Yang-Mills, magnetic or electric fields~\cite{naka} or vorticity ($\nabla \times \mathcal{A}_{\mu,\nu}$) in fluid dynamics. If we now define the wavefunction as a plane wave~\cite{tknn}, $\psi_J = u_{J(|\textbf{k}|_1,|\textbf{k}|_2)} = \varphi_{J(|\textbf{k}|_1,|\textbf{k}|_2)}\exp(-{\rm i}|\textbf{k}|_1x -{\rm i}|\textbf{k}|_2y) = u_{J(|\textbf{k}|^{\mu},|\textbf{k}|^{\nu})} = \varphi_{J(|\textbf{k}|^{\mu},|\textbf{k}|^{\nu})}\exp(-{\rm i}|\textbf{k}|^{\mu}x^{\mu} -{\rm i}|\textbf{k}|^{\nu}x^{\nu})$, and after using Eqs.~(\ref{eq:11}),~(\ref{eq:14}) and~(\ref{eq:18}), one obtains,
\begin {eqnarray}
G^{\rm TKNN}_{\rm Hall} = \frac{{\rm i}e^2}{2\pi h}\sum_J^L\int_{\texttt{S}({\rm d}|\textbf{k}|^{\mu}\wedge {\rm d}|\textbf{k}|^{\nu})}\int_{\texttt{S}({\rm d}x^{\mu}\wedge {\rm d}x^{\nu})}\big[\nabla \times \mathcal{A}^{J}_{j(\mu,\nu)}\big], \label{eq:19}
\end {eqnarray}  
which is nothing but the Hall conductance derived by Thouless et al.~\cite{tknn}, which is also known as the Thouless-Kohmoto-Nightingale-den Nijs (TKNN) equation. Here, $\int_{\texttt{S}({\rm d}|\textbf{k}|^{\mu}\wedge {\rm d}|\textbf{k}|^{\nu})}$ and $\int_{\texttt{S}({\rm d}x^{\mu}\wedge {\rm d}x^{\nu})}$ integrate the Fermi surface in momentum and real spaces, respectively, while $\sum_J^L$ sums all the occupied bands ($u_J, \cdots, u_L$) that contribute to the conductance. On the other hand, if we were to use Eqs.~(\ref{eq:16a}) and~(\ref{eq:18a}), then  
\begin {eqnarray}
G_{\rm Hall} &=& -\frac{{\rm i}e^2}{2\pi h}\int_{\texttt{S}}\big[\mathcal{F}^{J}_{k(\mu\nu)} - \mathcal{F}^{J}_{j(\mu\nu)}\big], \label{eq:20} \\&=& -\frac{e^2}{h}n_{\rm Chern} = G^{\rm Kohmoto}_{\rm Hall}. \label{eq:21}
\end {eqnarray}  
Here, Eq.~(\ref{eq:21}) is exactly the conductance derived by Kohmoto in Ref.~\cite{koh} where $J$ is the $J^{\rm th}$ conduction band. However, the origin of the negative sign that appears in Eq.~(\ref{eq:20}) has got nothing to do with the negative sign in the Kohmoto conductance formula, $G^{\rm Kohmoto}_{\rm Hall}$ derived in Ref.~\cite{koh}, in which, the source of the negative sign in $G^{\rm Kohmoto}_{\rm Hall}$ is i$^2$. In fact, we can reversibly switch the sign ($- \longleftrightarrow +$) in Eq.~(\ref{eq:20}), such that, the positive sign is for $\texttt{t}_{jk}$, while $\texttt{t}_{kj} = \texttt{t}^{-1}_{jk}$ implies a negative conductance, or vice versa, which then, can be defined to be related to a hole- or an electron-conduction. This reversible sign-switch is also compatible with $G^{\rm Kohmoto}_{\rm Hall}$ given in Ref.~\cite{koh} if one were to use the Chern winding number in the form of Eq.~(\ref{eq:16}). 

In summary, we have provided an alternative derivation to obtain the integer Hall conductance by explicitly invoking the definition for Hall conductance ($e^2n_{\rm Chern}/h$), the PB phase ($\gamma$) and the first Chern winding number ($c_1(\mathcal{F}) = n_{\rm Chern}$). You should be aware here that Eq.~(\ref{eq:21}) is valid for a two-dimensional (2D) Fermi gas or Fermi liquid metals such that $E_{\rm gap} = 0$, and this means that the Fermi-Dirac probability always equals one such that an electron can readily occupy an empty energy level at the Fermi surface without any energy penalty. 

\section{Hall resistance with irrelevant and finite energy-level spacings}

The Chern winding number introduced above actually refers to a winding fibre~\cite{naka}, a mathematical notion that has got nothing to do with any physical entity nor refers to an electron's energy level and this fibre does not represent the electronic wavefunction. For example, in the presence of external magnetic fields, $n_{\rm Chern}$ does not count the number of complete circles (multiple of 2$\pi$) made by an electron (or a hole) during the Hall transport. This means that, a particular wavefunction picks up a phase factor (given in Eq.~(\ref{eq:1})) on the right-hand side of a wavefunction (see Eq.~(\ref{eq:10a})), not because of the right-hand action operator (defined in Section 10.1 in Ref.~\cite{naka}), but due to a physical notion, known as the Pancharatnam phase retardation~\cite{panch}. This phase retardation originates from the interaction between an electron and the vector potential giving rise to the changing phase and/or group momenta~\cite{panch3}.

Rightly so, $n_{\rm Chern}$ has been replaced by the Laughlin-Jain filling factor, $\nu_{\rm LJ}$ worked out by Laughlin by means of the trial-wavefunction approach~\cite{laugh} and Jain~\cite{jain} such that $\nu_{\rm LJ}$ does not have to be an integer due to the interaction between charge carriers and the magnetic fields, $\textbf{B} = \nabla \times \textbf{A}$ where $\textbf{A}$ is a vector potential. In this case, Eq.~(\ref{eq:21}) reads, $G^{\rm LJ}_{\rm Hall} = -e^2\nu_{\rm LJ}/h$. Moreover, according to Jain~\cite{jkjain}, the integer $\nu_{\rm LJ}$ refers to composite fermions. The readers can refer to an excellent review by Murthy and Shankar~\cite{mur} on quantum Hall effects for $\nu_{\rm LJ} < 1/2$ within the extended Hamiltonian theories. There is also an alternative interpretation due to Wilczek~\cite{wil} where the fractional $\nu_{\rm LJ}$ is proposed to arise from the exotic particles called anyons.  

However, the game now has changed due to the rediscovery of Pancharatnam's phase retardation notion~\cite{panch3}, in such a way that one can substitute $n_{\rm Chern}$ with $\nu_{\rm P}$ that will capture the changes in the phase and group momenta of the electrons, in the presence of (i) $\textbf{B} \neq 0$ and (ii) irrelevant and finite energy-level spacings. We will see why this is so shortly. We first show why the irrelevant energy-level spacing, $\xi_{\rm irr}$, exists even in the presence of crossed energy levels such that the band and Mott-Hubbard energy gaps are zero. In this case, the electrons still satisfy the metallic Fermi liquid theory in the usual sense. For the free-electron metals, $\xi = 0$, and on the other hand, one obtains a strange metal if $\xi$ is a relevant parameter~\cite{pc}. When $\xi$ is zero or irrelevant, one can define the wavefunction as a plane wave because $\textbf{k}$ is ``energetically'' continuous ($E(\textbf{k})$ or the Fermi surface is continuous) throughout the momentum space, even within the so-called conduction band (with $E_{\rm gap} \neq 0$) or the overlapped band (with $E_{\rm gap} = 0$). This means that, even when $\xi = \xi_{\rm irr} \neq 0$, there is no energy gap for an electron to occupy the Landau level, $E_b(\textbf{k}_1)$ from its initial level, $E_a(\textbf{k}_1)$, hence the label, irrelevant in $\xi_{\rm irr}$ (see the discussion below).

Having said that, $\xi_{\rm irr}$ can be formally shown to exist from Refs.~\cite{pc,ptp},
\begin {eqnarray}
H(\textbf{k})\varphi_a(\textbf{k}) &=& [H_0(\textbf{k}) + \mathcal{V}(\textbf{k})]\varphi_a(\textbf{k}) \nonumber \\&=& [h_a(\textbf{k}) + \texttt{v}_a(\textbf{k})]\varphi_a(\textbf{k}) = E_a(\textbf{k})\varphi_a(\textbf{k}), \label{eq:22}
\end {eqnarray}
\begin {eqnarray}
H(\textbf{k})\varphi_b(\textbf{k}) &=& [H_0(\textbf{k}) + \mathcal{V}(\textbf{k})]\varphi_b(\textbf{k}) \nonumber \\&=& [h_b(\textbf{k}) + \texttt{v}_b(\textbf{k})]\varphi_b(\textbf{k}) = E_b(\textbf{k})\varphi_b(\textbf{k}). \label{eq:23} 
\end {eqnarray}
where $H(\textbf{k})$ is a solved Landau two-level ($\varphi_a(\textbf{k})$ and $\varphi_b(\textbf{k})$) Hamiltonian. In particular, $H_0(\textbf{k})$ denotes the non-interacting Hamiltonian and $\mathcal{V}(\textbf{k})$ is the interaction operator. The Landau energy levels, $E_a(\textbf{k}_1) = E_b(\textbf{k}_1)$ at a certain $\textbf{k}$ point ($\textbf{k}_1$) does not imply $\xi(\textbf{k}_1) = 0$ even though $E_{\rm gap}(\textbf{k}_1) = E_a(\textbf{k}_1) - E_b(\textbf{k}_1) = 0$ because $h_a(\textbf{k}_1) \neq h_b(\textbf{k}_1)$ and $\texttt{v}_a(\textbf{k}_1) \neq \texttt{v}_b(\textbf{k}_1)$. These not-equal signs and $E_{\rm gap}(\textbf{k}_1) = 0$ mean that the Landau levels are degenerate and
\begin {eqnarray}
&&\xi(\textbf{k}) = \sum_{i}E_a(\textbf{k}_i) - \sum_{i}E_b(\textbf{k}_i) \neq 0, \label{eq:24} \\&& \sum_{i}E_a(\textbf{k}_i) < \sum_{i}E_b(\textbf{k}_i), \label{eq:25}   
\end {eqnarray}
where $\xi_{\rm irr}$ is the averaged irrelevant LL spacing. If $\xi(\textbf{k}) = \xi_{\rm irr} \neq 0$, then the only required wavefunction transformation refers to picking up or dropping the PB phase factor (Eq.~(\ref{eq:14})) such that, the phase and group momenta, as well as the degenerate Landau levels (and the filling factor, $\nu_{\rm P}$) change with increasing or decreasing magnetic fields. Here, the changes to the PB phase factor is large (because the electron-flow involves many LLs) such that there is a large change in the group momentum of a wavefunction. On the other hand, $\xi(\textbf{k}) = \xi_{\rm irr} = 0$ restricts the wavefunction to pick up or drop the same PB phase factor within the same Landau level where the new filling factor, $\nu_{\rm P}$ is a constant, even when $B$ changes. In this case however, the changes to the PB phase factor require a small change in the group momentum of a wavefunction (because the electron-flow involves only one LL). 

As a consequence, the Hall resistance consists of two independent physical processes, which are activated when $\textbf{B}$ increases where changing $\textbf{B}$ initiates the changes to the Landau level crossings and Zeeman splittings, in such a way that there are two possibilities---(i) for a certain set of crossed Landau levels (within the degenerate LLs), one requires the electrons to flow within the same Landau levels (where $\nu_{\rm P}$ remains unchanged) governed by Eq.~(\ref{eq:21}). On the other hand, (ii) the second process is also activated for a different set of Landau level crossings where they now require the electrons to flow from one Landau level to another degenerate Landau level that gives ${\rm d}R_{\bot}/{\rm d}B \neq 0$. In this latter case, $\nu_{\rm P}$ is not a constant. Therefore, the total Hall resistance for quantum Hall metals can be constructed to read (we can suppress the negative sign for an obvious reason), 
\begin {eqnarray}
R_{\rm Hall} = \sum_q\bigg[\frac{h}{e^2}(\nu_{\rm P})_q\bigg]^{\rm Same}_{\rm LL,~\xi_{\rm irr} = 0} + \alpha_B\sum_r\bigg[\frac{{\rm d}R_{\bot}}{{\rm d}B}\bigg]_rB\bigg|^{\rm Many}_{\rm LLs,~\xi_{\rm irr} \neq 0}. \label{eq:26}
\end {eqnarray}
Here, the magnitude of applied magnetic fields is denoted by $B$, $\alpha_B$ is a numerical constant of proportionality due to applied magnetic fields, and recall that $\xi_{\rm irr} = 0$ is not literally true, but a symbolic way of saying that only one particular LL contributes to $R_{\rm Hall}$ (by means of the first term in Eq.~(\ref{eq:26})), which eventually means, $\xi_{\rm irr} \neq 0$ does not play any role. In particular, the first term in Eq.~(\ref{eq:26}) is independent of $\textbf{B}$ (because ${\rm d}R_{\bot}/{\rm d}B \approx 0$), and comes from the resistance due to electron-flow in the same Landau level (LL), labeled $q$. Whereas, the second term arises from the electron-flow involving many LLs within the degenerate LLs, labeled $r$, which is proportional to the change in the longitudinal resistance ($R_{\bot}$) along the applied electric fields, perpendicular to $R_{\rm Hall}$. For example, $R_{\rm Hall} \propto {\rm d}R_{\bot}/{\rm d}B$ because $R_{\bot}$ is not entirely due to applied electric fields if $\textbf{B} \neq 0$ where ${\rm d}R_{\bot}/{\rm d}B$ can be nonlinear. Besides that, whenever $q = r$, ${\rm d}R_{\bot}/{\rm d}B \approx 0$ where $r \neq q$ represent the electron's path with many $\nu_{\rm P}$ (due to many LLs or many quantitatively different wavefunctions are involved), while $q$ requires a constant $\nu_{\rm P}$ (due to a single LL or only an identical wavefunction is involved). Apparently, ``quantitatively different wavefunctions'' here means that they are orthonormalized, satisfying Eqs.~(\ref{eq:22}),~(\ref{eq:23}) and~(\ref{eq:24}). 
\begin{figure}
\begin{center}
\scalebox{0.26}{\includegraphics{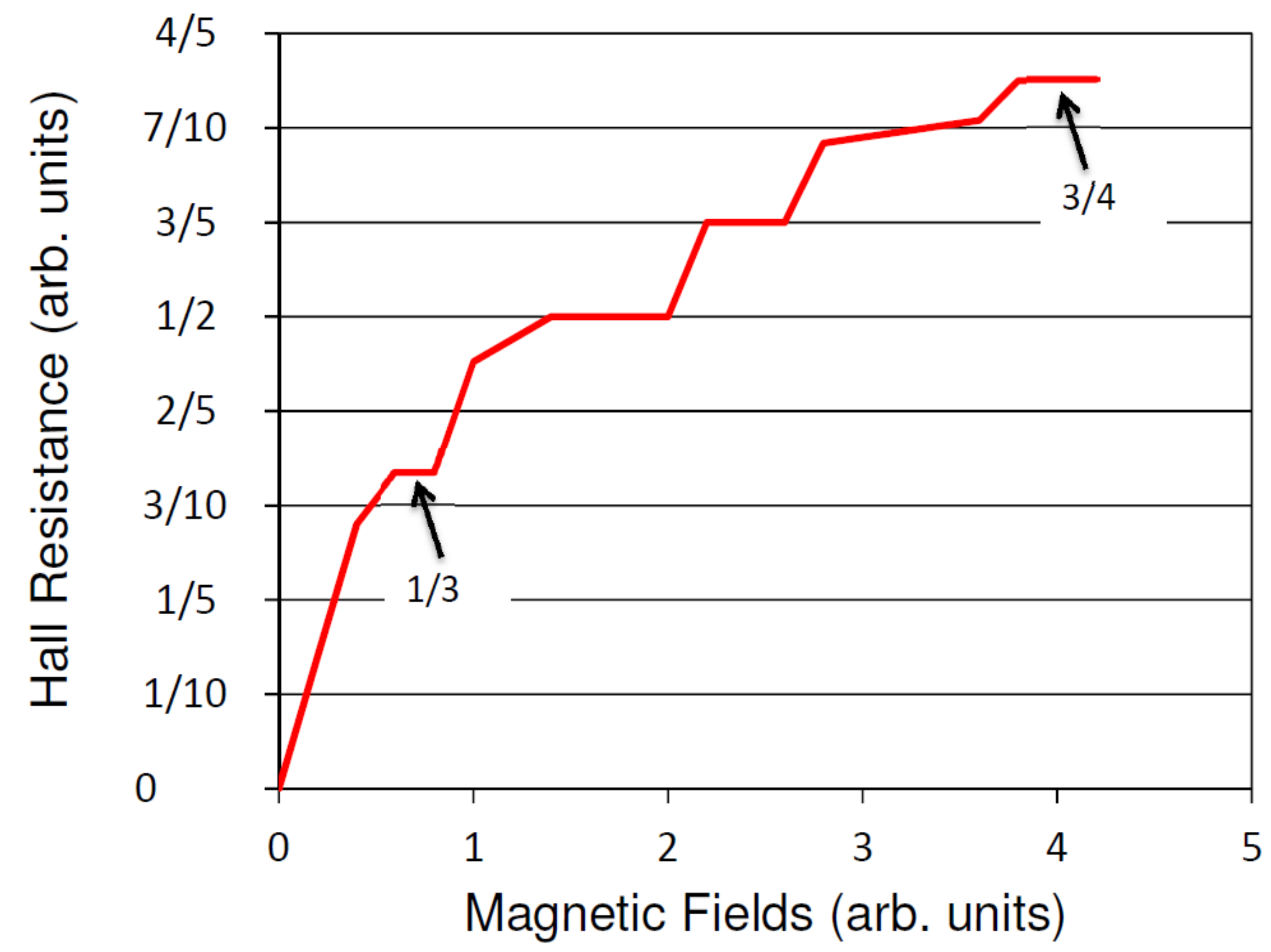}}
\caption{Calculated Hall resistance using Eq.~(\ref{eq:26}). The constant $R_{\rm Hall}$ (independent of the magnetic-field strength) is set to occur for $\nu_{\rm P} = \{1/3, 1/2, 3/5, 3/4,\cdots\}$ contributed by the first term in Eq.~(\ref{eq:26}), while the second term is by definition zero because in these cases, $q = r$. For $r \neq q$, the electron conduction involves many Landau levels (or many $\nu_{\rm P}$), and therefore, the second term in Eq.~(\ref{eq:26}) also contributes to the total $R_{\rm Hall}$ as a result of nonzero longitudinal resistance ($R_{\bot}$).}
\label{fig:1}
\end{center}
\end{figure}
In Fig.~\ref{fig:1}, we used Eq.~(\ref{eq:26}) to plot the zigzag Hall resistance curve by approximating $\alpha_B{\rm d}R_{\bot}/{\rm d}B = \alpha_B\Delta R_{\bot}/2\Delta B =$ $\alpha_B\times$slope and $h/e^2 = 1$ where the extra factor, the one-half that appears in the stated equality implies that we can approximately and correctly consider only the part where $R_{\bot}$ increases with increasing $B$, which then contributes to the total $R_{\rm Hall}$.

\section{Conclusions}

We have made use of the connection between the Pancharatnam phase retardation (changing phase and/or group momenta) and $R_{\rm Hall}$ via $\nu_{\rm P}$, $\xi_{\rm irr} = 0$ and $\xi_{\rm irr} \neq 0$ to construct Eq.~(\ref{eq:26}). Add to that, the physics that have led us to write the trial $R_{\rm Hall}$ given in Eq.~(\ref{eq:26}) do not require any anyons nor composite fermions where $\nu_{\rm P} \neq 1/\nu_{\rm LJ}$ because $\nu_{\rm P}$ can acquire any integer or fractions with odd or even denominators because with increasing $\textbf{B}$ (or other valid external disturbances, namely, the gate potential (${\rm d}R_{\bot}/{\rm d}V_{\rm gate}$) or external pressure (${\rm d}R_{\bot}/{\rm d}P$)) one can initiate the complicated changes to the LL crossings and Zeeman splittings (this is only true for $\textbf{B} \neq 0$), which in turn give rise to an effectively random LL crossings and $\nu_{\rm P}$. This also means that, the trial $R_{\rm Hall}$ and its physics can be proven to be false with proper Hall measurements for different types of samples. For example, if the integer and fractional values for $\nu_{\rm LJ}$ can be shown to be fundamental such that the same values of $\nu_{\rm LJ}$ (but for different sets of $R_{\rm Hall}$ and $\textbf{B}$) can also be systematically observed for other quantum Hall metals, then our physical postulates are definitely incorrect.

\section*{Acknowledgments}

This work was supported by Sebastiammal Innasimuthu, Arulsamy Innasimuthu, Amelia Das Anthony, Malcolm Anandraj and Kingston Kisshenraj. I am grateful to Alexander Jeffrey Hinde (The University of Sydney) and Marco Fronzi (National Institute for Materials Science, Japan) for their help in providing some of the references.

\end{document}